\documentclass[aip,reprint,jcp]{revtex4-1}
\usepackage{graphicx}
\usepackage{amsmath,amssymb}
\usepackage{color}
\usepackage{float}
\newcommand{\revise}[1]{\textcolor{black}{#1}}
\begin{document}

\title{The nature of the hydrophobic interaction varies as the solute size increases from methane's to C$_{60}$'s}

\author{Hidefumi Naito}
\author{Tomonari Sumi}
\author{Kenichiro Koga}
\email[Author to whom correspondence should be addressed: ]{koga@okayama-u.ac.jp}
\affiliation{Department of Chemistry, Faculty of Science, Okayama University, Okayama 700-8530, Japan}
\affiliation{Research Institute for Interdisciplinary Science, Okayama University, Okayama 700-8530, Japan}

\date{\today}

\begin{abstract}
The hydrophobic interaction, often combined with the hydrophilic or ionic 
interactions, makes the behavior of aqueous solutions very rich and 
plays an important role in biological systems. Theoretical and computer simulation studies haven shown that the water-mediated force depends strongly on the size and other chemical properties of the solute, but how it changes with these factors remains unclear. We report here a computer simulation study that illustrates how the hydrophobic pair interaction and the entropic and enthalpic terms change with the solute size when the solute-solvent weak attractive interaction is unchanged with the solute size. The nature of the hydrophobic interaction changes qualitatively as the solute size increases from that of methane to that of fullerene. The potential of mean force between small solutes has several well-defined extrema including the third minimum whereas the potential of mean force between large solutes has the deep contact minimum and the large free-energy barrier between the contact and the water-bilayer separated configurations. The difference in the potential of mean force is related to the differences in the water density, energy, and hydrogen bond number distributions in the vicinity of the pairs of hydrophobic solutes.
\end{abstract}


\maketitle

\section{Introduction}

\noindent
The effective interactions between solute molecules in water or in aqueous solutions in many cases differ qualitatively from the corresponding direct interactions in vacuum due to the solvent-induced part of the interactions. 
More important than the mere difference from the direct interaction, however, is that the effective interaction depends in a complex way on the temperature, pressure and composition of an aqueous solution, as well as the size of the solute and the solute-solvent interaction. 

The hydrophobic interaction is one of the most studied effective interactions as it is the driving force for molecular self-assembly, such as the formation of micelles, membranes, and vesicles in aqueous solutions.\cite{Kauzmann1959, tanford1980, ben-naim1980, doi:10.1021/cr000692+,southall2002jpcb, widom2003pccp} 
Small hydrophobic molecules such as methane and inert gases are less attracted (or even repelled) in water than in a vacuum at low temperatures, but more attracted than in a vacuum at high temperatures.~\cite{koga2013jpcb} The pressure and salt-concentration dependences
of the hydrophobic interaction are also similar in trend to the temperature dependence.\cite{koga2018jpcb} 
The character of the hydrophobic hydration changes with the size of the solute,\cite{chandler2005nature} and so would the nature of the hydrophobic interaction, including its temperature, pressure, and salt concentration dependencies.

In fact, the effect of the solute size on the strength of the hydrophobic interaction has been one of the issues that has attracted attention.~\cite{pratt1977jcp,lum1999,tanaka1987jcp,hummer1996pnas,southall2002biochem,sumi2007jcp,sobolewski2007jpcb,makowski2010jpcb,bartosik2015jpoc,bogunia2020jpcb,graziano2009jpcb,graziano2010cpl,graziano2017cpl,zangi2011jpcb,BenAmotz2015jpcl,BenAmotz2016arpc,sun2019chemphys,naito2022jcp,naito2024fd}
Lum, Chandler, and Weeks showed that the solvation free energy and the effective interaction of hard-sphere solutes in water exhibit the crossover occurring on nanometer length scales.~\cite{lum1999}
Zangi quantified the strength of hydrophobic interaction between graphene sheets with different sizes, showing that the association of small graphene sheets has an entropic origin while 
that of large graphene sheets is mainly due to the solute-solute direct pair interaction.~\cite{zangi2011jpcb}
Scheraga et al. calculated the potential $w(r)$ of the mean force for hydrocarbons of different sizes in water,~\cite{sobolewski2007jpcb,makowski2010jpcb} the results of which indicate that $w(r)$ at the contact distance becomes higher than the potential in vacuum as the size of the solute exceeds that of neopentane.
In a broader context, earlier studies based on the theory of liquids have examined 
the solute-size dependence of the solvent-mediated interactions in simple liquids.~\cite{sumi2007jcp,kimura1991MP,attard1989jcp,attard1990jcp,dickman1997jcp,biben1996jpcm,dijkstra1999pre,roth2000pre,roth2006jcp,kinoshita2002cpl,akiyama2006jpcj,simonin2001jpcb}
Numerical results from the integral equation method~\cite{kimura1991MP} 
and molecular dynamics (MD) simulations~\cite{naito2022jcp, naito2024fd} indicate that the effective pair interactions between Lennard-Jones (LJ) particles in LJ liquids become less attractive 
with the size of the solute, provided that the LJ energy parameter for 
the solute-solvent pair is fixed to a value greater than or equal to the LJ energy 
parameter for the solvent-solvent pair.~\cite{naito2022jcp,naito2024fd,kimura1991MP} 


The strength of the effective solute-solute pair interactions in a solvent can be quantified by the osmotic second virial coefficient $B$ in the virial expansion of the osmotic pressure at constant chemical potential of the solvent species. The effective interaction is overall attractive when $B$ is negative and repulsive otherwise.
$B$ is related to $w(r)$ or the solute-solute radial distribution function $g(r)$ via 
\begin{align}
\label{eq_B}
B&=-\frac12\lim_{\rho\rightarrow0}\int\Bigl[\exp\Bigl[-\frac{w(r)}{kT}\Bigr]-1\Bigr]d\tau\nonumber\\
&=-\frac12\lim_{\rho\rightarrow0}\int[g(r)-1]d\tau,
\end{align}
where $\rho$ is the solute density, $k$ is  Boltzmann's constant, $T$ is the temperature, 
and $d\tau$ is an infinitesimal volume element.~\cite{mcmillan1945jcp} 
Recent simulation studies have found that $B$, which is negative, is proportional to some power $\alpha$ of the solute diameter, and that $\alpha \simeq 6$ or 7, indicating 
an extremely strong size dependence for LJ solutes in water.~\cite{naito2022jcp,naito2024fd}

Aqueous solutions of water-soluble polymers such as PNIPAM and PEO are homogeneous at room temperature but undergo phase separation at higher temperatures, indicating that the effective interactions between hydrophobic moieties in those polymers are entropic forces.  
It has been well verified by simulations and theoretical methods that the strength of the hydrophobic interaction increases with temperature.~\cite{pratt1977jcp,smith1992jacs,smith1993jcp,jungwirth1994cpl,rick1997jpcb,rick2003jpcb,shimizu2000jcp,shimizu2001jacs,paschek2004jcp1,paschek2004jcp2,graziano2009jpcb,graziano2016jpcj,zieba2020pccp,bogunia2022jpcb,graziano2017cpl,chaudhari2013pnas,Pratt:2016hm,chaudhari2016jpcb,koga2013jpcb,Ashbaugh:2015cx,tang2018jcp,bartosik2015jpoc,cerdeirina2016jpcb,koga2018jpcb,naito2022jcp} 
Although the hydrophobic interaction is overall an entropic force, but there is a significant  enthalpic (energetic) contribution, which is opposite in sign to the entropic one.
~\cite{smith1992jacs,smith1993jcp,jungwirth1994cpl,ludemann1997jacs,rick1997jpcb,rick2000jpcb,rick2003jpcb,islam2019jcp,shimizu2000jcp,shimizu2001jacs,southall2002biochem,paschek2004jcp1,paschek2004jcp2,zangi2008jpcb,zangi2011jpcb,graziano2009jpcb,graziano2016jpcj,graziano2017cpl,zieba2020pccp,sobolewski2012jpcb,bartosik2015jpoc,bogunia2022jpcb,Ashbaugh:2015cx}
To fully understand the solute size dependence of $w(r)$, it is essential to investigate how the fraction of the entropic contribution to $w(r)$ changes with solute size. 

The aim of the present work is to quantify the enthalpic and entropic contributions to $w(r)$ for solutes of different sizes, and thereby to reveal how the underlying molecular mechanisms in the hydrophobic interaction change with solute size.
First, we calculate with high accuracy $w(r)$ and $B$ for the LJ particles with different diameters $\sigma$ in water at several temperatures. The enthalpic and entropic contributions to $w(r)$ are obtained from the temperature derivative of the solvent-induced part $w^*(r)$.
Second, we analyze the microscopic structures of water molecules around two solutes when the solute-solute distance $r$ is fixed to some characteristic distances.

\section{Computational Details}

\noindent
Isobaric-isothermal molecular dynamics (MD) simulations were performed for the model aqueous solutions consisting of TIP4P/2005 water molecules~\cite{abascal2005jcp} and spherical solute particles interacting via the Lennard-Jones (LJ) potential:
\begin{equation}
\label{eq_LJ}
\phi(r)=4\epsilon\biggl[\Bigl(\frac{\sigma}{r}\Bigr)^{12}-\Bigl(\frac{\sigma}{r}\Bigr)^{6}\biggr],
\end{equation}
where $\epsilon$ is the depth of the pair potential well and $\sigma$ is the particle diameter.
The reference solute is chosen to be methane, whose LJ parameters are  $\sigma_{\mathrm{m}}=0.373{\mathrm{\ nm}}$ and $\epsilon_{\mathrm{m}}=1.23{\mathrm{\ kJ \ mol^{-1}}}$ in the TraPPE-UA force field.~\cite{martin1998jpcb} 
The reduced LJ diameters for solutes are $\sigma^*\equiv\sigma/\sigma_{\mathrm{m}}=1, 1.5, 2, 2.5, {\mathrm{\ and \ }} 3$; the solute-solute LJ energy parameter $\epsilon$ is fixed to $\epsilon_{\mathrm{m}}$.
The solute-water pair potential is the LJ potential whose size and energy parameters are given by the Lorentz-Berthelot combining rules.

We used GROMACS 2018 software~\cite{abraham2015softwarex} to perform MD simulations for the aqueous solutions under three-dimensional periodic boundary conditions. 
The pressure is set to 1 bar using the Parrinello-Rahman method and 
the temperature is fixed at 270, 300, 330, or 360~K by the Nosé-Hoover method.
The equilibrium trajectories for $T=300{\mathrm{\ K}}$ are those obtained earlier.~\cite{naito2024fd} 
The time step interval is 1 fs and the configurations of 
solute particles were recorded every 0.05 ps.


The model aqueous solutions consist of water molecules and solute particles.
For the solutes with smaller diameters $\sigma^*=$ 1 and 1.5, 
the number $N$ of solute particles is 40 (for $\sigma^*=1$) or 20 (for $\sigma^*=1.5$)
and the number $N_{\rm w}$ of water molecules is 4000. The potential $w(r)$ of mean force 
for each system was computed via the solute-solute $g(r)$ obtained from the MD trajectories. 
The duration time $t$ for the production run is 100 ns for $\sigma^*=1$ and 200 ns for $\sigma^*=1.5$.
For the solutes with larger diameters $\sigma^*=$ 2, 2.5, and 3,  
$N=2$ and $N_{\rm w}=8000$. For these systems, we performed the umbrella sampling simulations~\cite{torrie1974cpl,torrie1977jcp} to compute $w(r)$.
The simulation time $t$ for each window is 20 ns. 


In the MD simulations for the solutes with $\sigma^*=$ 1 and 1.5, the solute-solute LJ pair potential $\phi(r)$ was replaced with a repulsive potential $\phi_{\mathrm{rep}}(r)$ 
in order to suppress multi-body aggregations and minimize the finite-concentration effect on $g(r)$. This procedure is particularly important to evaluate the osmotic second virial coefficient $B$ from Eq.~\eqref{eq_B}.
The resulting radial distribution function $g_{\mathrm{rep}}(r)$ is then converted to $g(r)$ by
\begin{equation}
\label{eq_finite_conc}
g(r)=g_{\mathrm{rep}}(r)\exp{\biggl[-\frac{\phi_{\mathrm{att}}(r)}{kT}\biggr]}
\end{equation}
with $\phi_{\mathrm{att}}(r)=\phi(r)-\phi_{\mathrm{rep}}(r)$.
For $\sigma^*=1$, $\phi_{\mathrm{rep}}(r)$ is the repulsive part of the Weeks-Chandler-Andersen (WCA) potential.~\cite{weeks1971jcp} The solute particles with $\sigma^*=1.5$ tend to aggregate in water more strongly than those with $\sigma^*=1$. Therefore, $\phi_{\mathrm{rep}}(r)$ is taken to be $4\epsilon(\sigma/r)^{12}$.

The cutoff distance $r_{\mathrm{cut}}$ for the LJ potentials depends on the diameter of the solute particle in each aqueous solution:
$r_{\mathrm{cut}}=1.3$ nm for $\sigma^*=1{\mathrm{\ and \ }}1.5$, $r_{\mathrm{cut}}=2$ nm for $\sigma^*=2{\mathrm{\ and \ }}2.5$, and $r_{\mathrm{cut}}=2.4$ nm for $\sigma^*=3$.
The Coulomb potentials were treated using the particle mesh Ewald method, with the real-space cutoff distance being the same as $r_{\mathrm{cut}}$ for the LJ potentials.

In the umbrella sampling simulations, the solute-solute separations were constrained via the harmonic potential with the spring constant set to $1000 {\mathrm{\ kJ}}$ $\mathrm{mol^{-1}}$ $\mathrm{nm^{-2}}$.
The constraint distance ranges from 0.6~nm to 2.9~nm in 0.1~nm increments for the solute with $\sigma^*=2$, from 0.7~nm to 2.8~nm for $\sigma^*=2.5$, and from 0.8~nm to 3.1~nm for $\sigma^*=3$.
The potentials $w(r)$ of mean force were obtained from the umbrella sampling simulations for 
the windows using the weighted histogram analysis method.~\cite{kumar1992jcc,souaille2001cpc}
The corresponding radial distribution functions $g(r)$ are given by $\exp{[-w(r)/kT]}$.

The osmotic second virial coefficient $B$ is evaluated from $g(r)$ via Eq.~\eqref{eq_B}.
However, one cannot use $g(r)$ directly obtained from a simulation since 
in a closed system it does not converge to 1 and the volume integral of $g(r)-1$ 
would tend to diverge. To overcome this issue we employed the following computational procedure.~\cite{naito2022jcp,naito2024fd} 
First, the entire function $g(r)$ was scaled by a factor adjusting the average value of $g(r)$ over 
a certain range of large $r$ to 1.~\cite{koga2013jpcb}
Second, the Kirkwood-Buff (KB) integral $G$ is evaluated from the following equations:~\cite{kruger2013jpcl,dawass2020nanomaterials}
\begin{equation}
\label{eq_kruger_1}
G(L)=\int_0^L[g(r)-1]4\pi r^2\biggl[1-\frac32\Bigl(\frac{r}{L}\Bigr)+\frac12\Bigl(\frac{r}{L}\Bigr)^3\biggr]dr,
\end{equation}
\begin{equation}
\label{eq_kruger_2}
G(L)L=GL+D,
\end{equation}
where $L$ is the upper limit of the integral with respect to $r$, and $D$ is a constant.
$G(L)$ is obtained as a function of $L$ using Eq.~\eqref{eq_kruger_1} with corrected $g(r)$.
$G$ is then determined as a slope in the plot of $G(L)L$ vs.~$L$ over a certain range where $G(L)L$ is linear to $L$. 

The solute-solute effective potential $w(r)$ consists of the direct part $\phi(r)$ and the indirect, solvent-induced part $w^*(r)=w(r)-\phi(r)$. 
The latter is the difference between the solvation free energy $\mu_\text{pair}^*(r)$ of a pair of solutes distance $r$ apart and that of a pair infinitely far apart.
The temperature dependence of $w(r)$ comes from that of $w^*(r)$ alone as $\phi(r)$ is independent of $T$. 
The temperature derivative of $w^*(r)$ at fixed pressure gives the enthalpic and entropic contributions to $w(r)$:
\begin{equation}
\label{eq_entropy_enthalpy}
\Delta h^*(r)=\bigg[\frac{\partial w^*(r)/T}{\partial 1/T}\biggr]_p,~~~~
\Delta s^*(r)=-\bigg[\frac{\partial w^*(r)}{\partial T}\biggr]_p,
\end{equation}
where $\Delta$ means the constant-pressure process of changing 
the distance between two solute particles from infinity to $r$;   
$h^*(r)$ and $s^*(r)$ are the solvation enthalpy and entropy of the pair at fixed pressure.
Here the solvation enthalpy and entropy at fixed pressure mean the enthalpy and entropy changes upon 
inserting a molecule (now a pair of solute particles) in the solvent at a given pressure with 
its position and orientation fixed.~\cite{koga2011pccp}

In the present study, we evaluated $w^*(r)$ at four temperatures (270 K, 300 K, 330 K, and 360 K) 
in order to fit the data to a quadratic function of $T$
\begin{equation}
\label{eq_T_quadratic}
w^*_\text{fit}(r) = T^2a(r)+Tb(r)+c(r).
\end{equation}
Then we evaluated $\Delta s^*(r)$ and $\Delta h^*(r)$ at $T=300$~K via 
\begin{align}
\label{eq08}
\Delta h^*(r)=-T^2a(r) + c(r),~~~~~
\Delta s^*(r) = -2Ta(r)-b(r).
\end{align}

We performed additional isobaric-isothermal MD simulations for the aqueous solutions containing a pair of LJ solutes with fixed separation distance to examine the microscopic structures of water molecules 
around the pair. The solute diameters are $\sigma^*=1, 2$, and 3. 
In those simulations, the solute-solute distance $R$ was fixed to the values corresponding to the first minimum, the first maximum, the second minimum, the second maximum, and the third minimum of $w(r)$.
Table~\ref{table_2} shows the values of $R$ for $\sigma^*=1, 2, {\mathrm{\ and \ }}3$.
The pressure and temperature are maintained at 1 bar and 300 K, respectively.
The duration time of the production run is 20 ns for each simulation, and the configurations of solute particles and water molecules were recorded every 1 ps.
We also investigate the solute with $\sigma^*=3$ interacting with water molecules via the repulsive WCA potential. 

\begin{table*}[t]
\caption{
The fixed solute-solute distances $R$ corresponding to the first minimum, the first maximum, the second minimum, the second maximum, and the third minimum of the potentials $w(r)$ of mean force for pairs of LJ solutes with $\sigma^*=1, 2, {\mathrm{\ and \ }}3$ at $T=300{\mathrm{\ K}}$.
}
\label{table_2}
\begin{tabular*}{1\textwidth}{@{\extracolsep{\fill}}cccccc}
\hline
 & & & $R$ / nm & & \\
$\sigma^*$ &  1st minimum & 1st maximum & 2nd minimum & 2nd maximum & 3rd minimum \\
\hline
1 & 0.388 & 0.562 & 0.706 & 0.874 & 1.026 \\
2 & 0.724 & 0.992 & 1.114 & 1.230 & 1.414 \\
3 & 1.032 & 1.400 & 1.506 & 1.570 & 1.764 \\
\hline
\end{tabular*}
\end{table*}

\begin{figure}[ht]
\centering
\includegraphics[scale=0.4]{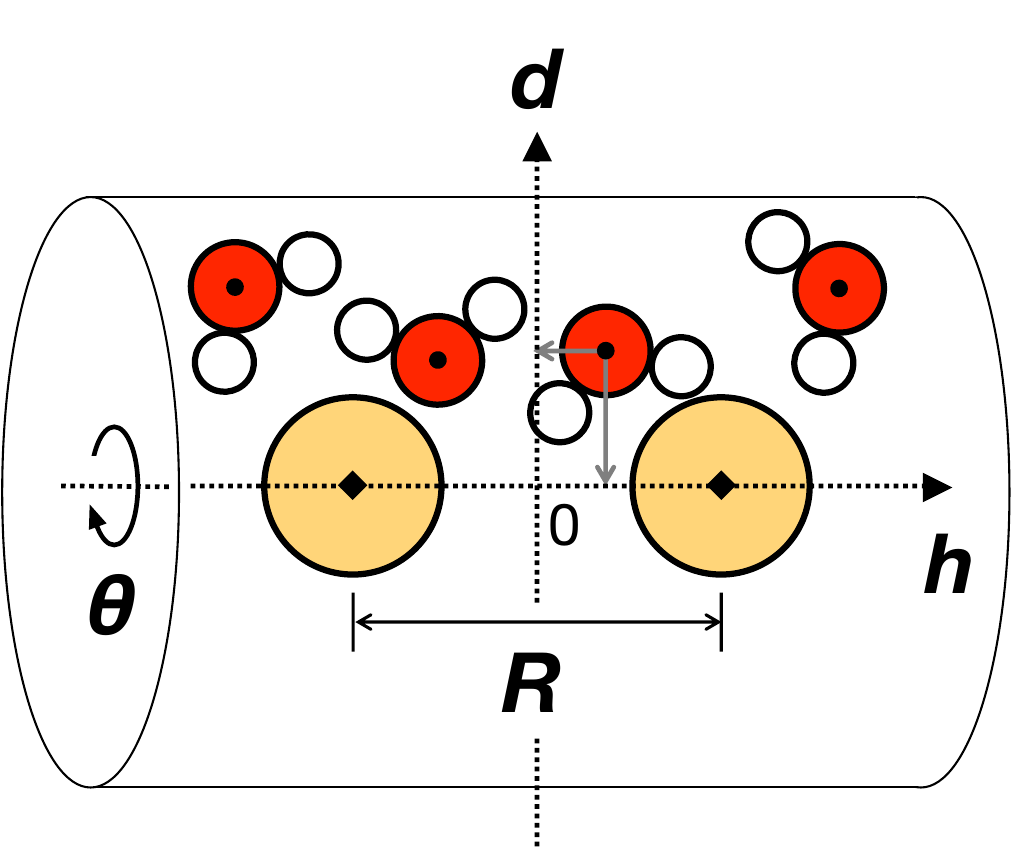}
\caption{
Schematic illustration of water molecules around two solute particles in the cylindrical coordinate system $(d, \theta, h)$.
The axis $h$ passes through two centers of solutes (orange circles), $d$ is the distance from the axis $h$, 
and $\theta$ is the azimuth. The origin of the coordinate system is set to 
the midpoint of the two centers of solutes. The red circles represent oxygen atoms of water molecules. 
The equilibrium distributions of oxygen atoms averaged over $\theta$ are depicted on 
the $h$-$d$ plane.  
}
\label{fig_cylindrical}
\end{figure}

Fig.~\ref{fig_cylindrical} is the schematic illustration of the cylindrical coordinate system ($d, \theta, h$).
Based on this system, we calculated the distribution $g_\text{cyl}(h,d)$ of water molecules (oxygen), which is normalized to unity in the bulk region, the water-water pair interaction energy $E_{\mathrm{ww}}(h,d)$, and the number $N_{\mathrm{HB}}(h,d)$ of hydrogen bonds, all the quantities being averaged over $\theta$ and being depicted on the $d$-$h$ plane. 
A pair of water molecules is taken into account for computing $E_{\mathrm{ww}}$ when the distance $r_{\mathrm{OO}}$ between their oxygen atoms is less than 0.35 nm.
Two water molecules form one hydrogen bond if $r_{\mathrm{OO}}<0.35{\mathrm{\ nm}}$ and the H-O$\cdots$O angle is less than or equal to 30 degrees.~\cite{luzar1996prl,soper1986cp,teixeira1990jpcm}
We note that the cylindrical distributions of water molecules around two solute particles and other quantities were evaluated in earlier studies.~\cite{southall2002biochem,paschek2004jcp2,sobolewski2007jpcb,makowski2010jpcb,zieba2020pccp,bogunia2020jpcb} Our focus here is on the question of how these distributions changes with the solute particle size.

\section{Results and Discussion}

\begin{figure*}[t]
\centering
\includegraphics[scale=0.43]{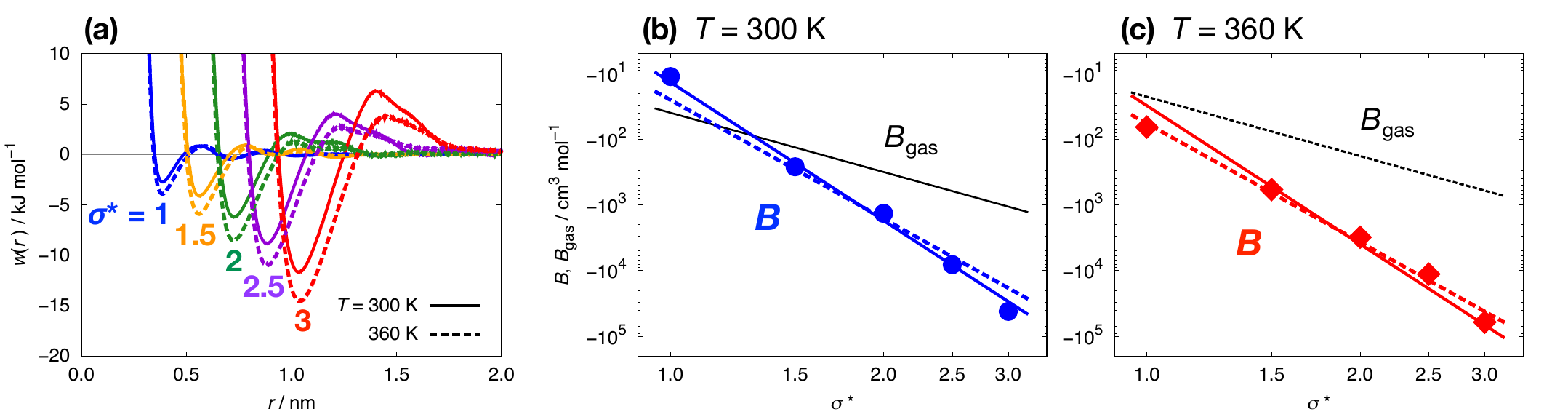}
\caption{
(a) The potentials $w(r)$ of mean force for pairs of solutes with various sizes in water at different temperatures $T$.
The LJ diameters of the solute particles are $\sigma^*\equiv\sigma/\sigma_{\mathrm{m}}=1,1.5,2,2.5,{\mathrm{\ and \ }}3$ with $\sigma_{\mathrm{m}}=0.373{\mathrm{\ nm}}$.
The solute-solute LJ energy parameter $\epsilon$ is fixed to $\epsilon_{\mathrm{m}}=1.23{\mathrm{\ kJ \ mol^{-1}}}$ for all-size solutes.
Solid and dotted curves are $w(r)$ at $T=300{\mathrm{\ K \ and \ }}360{\mathrm{\ K}}$, respectively.
(b) The log-log plot of the osmotic second virial coefficients $B$ at $T=300{\mathrm{\ K}}$ against $\sigma^*$.
The dotted and solid blue lines are the best fits to the data with the 6th and 7th power law, respectively.
The second virial coefficient $B_{\mathrm{gas}}$ for the LJ gas is also plotted as a function of $\sigma^*$.
(c) $B$ and $B_{\mathrm{gas}}$ at $T=360{\mathrm{\ K}}$.
The best fits of the 6th (dotted red line) and 7th (solid red line) power law are also plotted.
}
\label{fig_pmf_and_B}
\end{figure*}

\subsection{Potential of mean force and osmotic second virial coefficient}
First, we examine the size and temperature effects on the solute-solute effective interactions in water.
Fig.~\ref{fig_pmf_and_B}(a) shows the potentials $w(r)$ of mean force for pairs of LJ particles with different diameters in water at $T=300{\mathrm{\ K \ and \ }}360{\mathrm{\ K}}$.
The reduced solute diameter $\sigma^*$ ranges from 1 (0.373~nm, methane size) to 3 (1.119~nm, $\mathrm{C_{60}}$ size). 
At both temperatures, the first minimum of $w(r)$ descends with increasing $\sigma^*$.
In contrast, both the first maximum and the second minimum of $w(r)$ ascend with $\sigma^*$, and the latter disappears at $\sigma^*$ greater than 2, being a shoulder.
These features at ambient temperature have been reported earlier~\cite{naito2024fd} and now are confirmed at 360~K. One also notices that the curve of $w(r)$ shifts downward with increasing $T$. 

{
Note that $w(r)$ for the C$_{60}$-sized solutes is minimal at the distance smaller than the LJ diameter $\sigma$ (Table I). This is because the water-induced attractive force between the C$_{60}$-sized solutes is so strong that it balances with the equally strong repulsive force 
due to the direct solute-solute LJ pair potential at the short distance.}

Fig.~\ref{fig_pmf_and_B}(b) is the plot of the osmotic second virial coefficients $B$ at $T=300{\mathrm{\ K}}$ as a function of $\sigma^*$ in a log-log scale.
The osmotic $B$ is negative, and that magnitude increases with increasing $\sigma$.
This result indicates that the effective interaction between solute particles in water is attractive and becomes stronger as the particle size increases.
The second virial coefficients $B_{\mathrm{gas}}$ for the LJ gas, which quantify the strength of the direct pair interaction in a vacuum, are also plotted in Fig.~\ref{fig_pmf_and_B}(b).
Both $B$ and $B_{\mathrm{gas}}$ decrease with $\sigma$, but the magnitude of $B$ is much greater than that of $B_{\mathrm{gas}}$ in the $\sigma^*$ range over 1.5.

The log-log plot of $B$ vs.~$\sigma^*$ exhibits a linear relationship,  indicating that 
\begin{equation}
\label{eq_B_sigma_power}
B\propto(\sigma^*)^{\alpha}.
\end{equation}
The best fits of the data to Eq.~\eqref{eq_B_sigma_power} with $\alpha=6{\mathrm{\ and \ }}7$ are plotted by the dotted and solid blue lines, respectively.
One may see that $\alpha=7$ fits the data better than 6.
The gas virial coefficient $B_{\mathrm{gas}}$ is proportional to the cubic of the particle diameter, so the contribution from the water-induced part $w^*(r)=w(r)-\phi(r)$ to $B$ has a stronger size dependence than that from the direct part $\phi(r)$.
Earlier studies~\cite{naito2022jcp,naito2024fd} showed that the power law between $B$ and $\sigma$ is partially understood based on the thermodynamic identity~\cite{widom2012jpcb,koga2015jpcb} for $B$.

Fig.~\ref{fig_pmf_and_B}(c) displays $B$ and $B_{\mathrm{gas}}$ at $T=360 {\mathrm{\ K}}$ as a function of $\sigma^*$.
At this temperature, the magnitude of $B$ is greater than that of $B_{\mathrm{gas}}$ for any size of solute particles. 
The log-log plot of $B$ vs.~$\sigma^*$ again suggests the power-law dependence of $B$ on $\sigma^*$. The best estimate of $\alpha$ at 360 K is 6 rather than 7. 

Comparison between the plots in Fig.~\ref{fig_pmf_and_B}(b) and 
those in Fig.~\ref{fig_pmf_and_B}(c) indicates that 
$B$ decreases (becomes more negative) with increasing $T$ for any given solute, i.e., 
the effective solute-solute pair attractive force between hydrophobic particles in water becomes stronger at higher temperatures, 
which is characteristic of the hydrophobic interaction. On the other hand,
$B_{\mathrm{gas}}$ increases with $T$, which is known for real gases 
and for model pair potentials including the LJ potential.

\begin{figure*}[ht]
\centering
\includegraphics[scale=0.094]{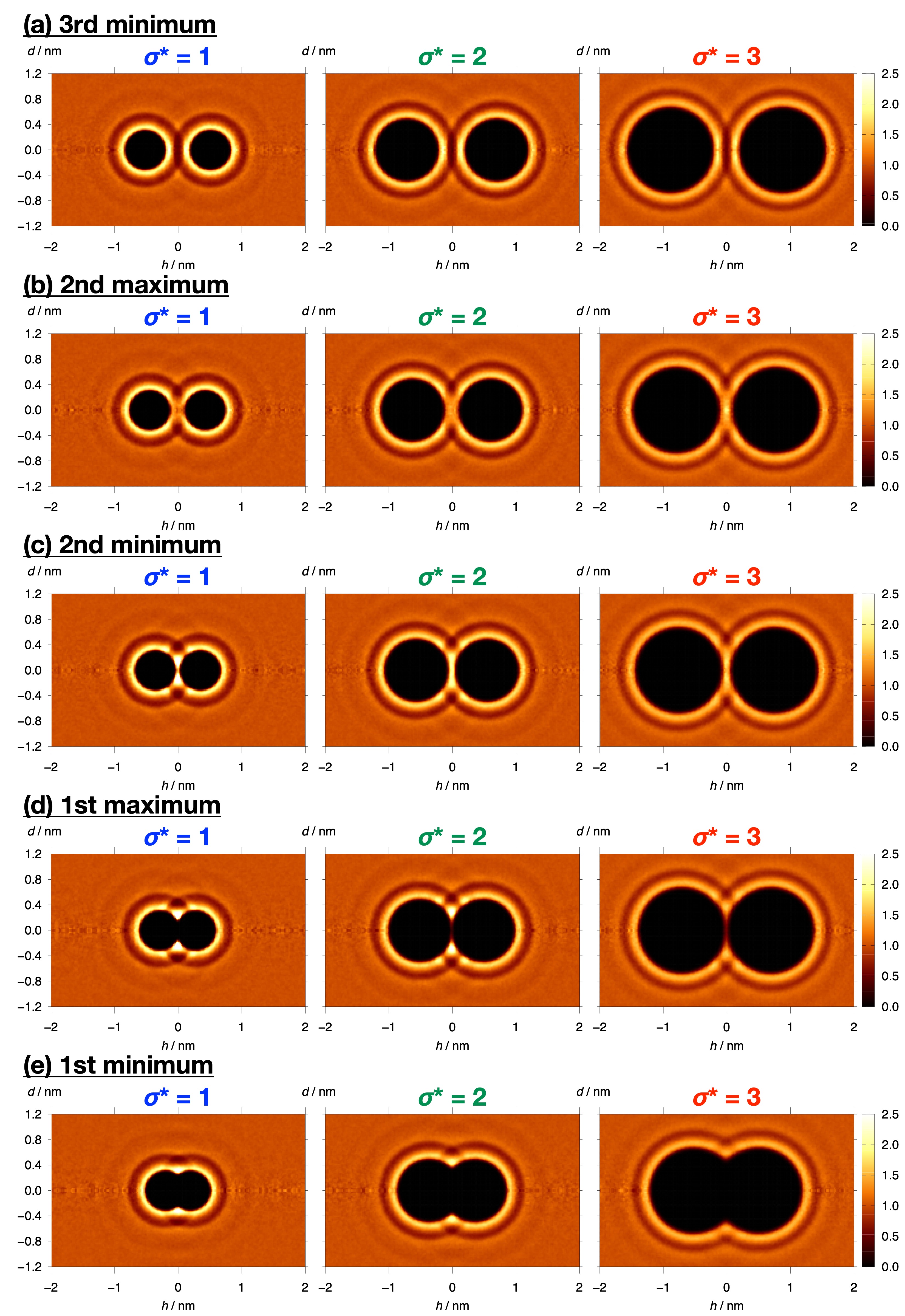}
\caption{
The normalized two-dimensional distributions $g_\text{cyl}(h,d)$ of water molecules around a pair of solutes at $T=300{\mathrm{\ K}}$.
Two solute particles with LJ diameters $\sigma^*=$ 1(left), 2(middle), and 3(right) are displayed in black.
The solute-solute separation distance $R$ is fixed to that of (a) the third minimum, (b) the second maximum, (c) the second minimum, (d) the first maximum, and (e) the first minimum of $w(r)$ for each solute (see Table~\ref{table_2}).
}
\label{fig_distribution}
\end{figure*}

\subsection{Distributions of water molecules around pairs of solute particles}
Now, we examine the distribution $g_\text{cyl}(h,d)$ of water molecules around a pair of solute particles with the inter-particle distance $r$ fixed to particular values. 
Fig.~\ref{fig_distribution}(a) shows $g_\text{cyl}(h,d)$ for $r$ of the third minimum of $w(r)$ (See Table~\ref{table_2} for the values of $r$). 
For all the diameters of solute particles, the first solvation shells (bright-colored rings) 
around the two particles have little effect on each other even in between the two particles. 
The local density of water at the solvation shell is much higher than the bulk density:
$g_\text{cyl}(h,d)$ is approximately 1.9, 1.8, and 1.6 for $\sigma^*=1$, 2, and 3, respectively.
At this seperation distance, $g_\text{cyl}(h,d)$ has two peaks in between the two solutes, indicating the bilayer structure of water.

The bilayer structure disappears at $r$ set to the second maximum of $w(r)$: the two solvation shells partly overlap with each other in the gap between 
the two spherical particles and $g_\text{cyl}$ in the gap is lower than outside for $\sigma^*=$ 1 and 2 (Fig.~\ref{fig_distribution}(b)). 
However, $g_\text{cyl}$ in the gap is higher than outside when 
the separation distance is fixed to that of the second minimum of $w(r)$ [Fig.~\ref{fig_distribution}(c)]: $g_\text{cyl}(0,0)$ is approximately 2.4 and 2.1 for $\sigma^*=1$ and 2. 

When the inter-particle distance is of the first maximum of $w(r)$ [Fig.~\ref{fig_distribution}(d)],  the distribution $g_\text{cyl}(0,0)$ at the midpoint of the pair is approximately 0, 0.4, and 0.6 for $\sigma^*=1$, 2 and 3, i.e., water molecules are scarce in the gap between the two particles. 
Finally, when the two solute particles are in contact [Fig.~\ref{fig_distribution}(e)], 
the distribution at the intersection of the two first solvation shells is greater than at the other part of the same solvation shells: See the brighter spots in the case of $\sigma^*=1$ and 2. 

\begin{figure*}[ht]
\centering
\includegraphics[scale=0.43]{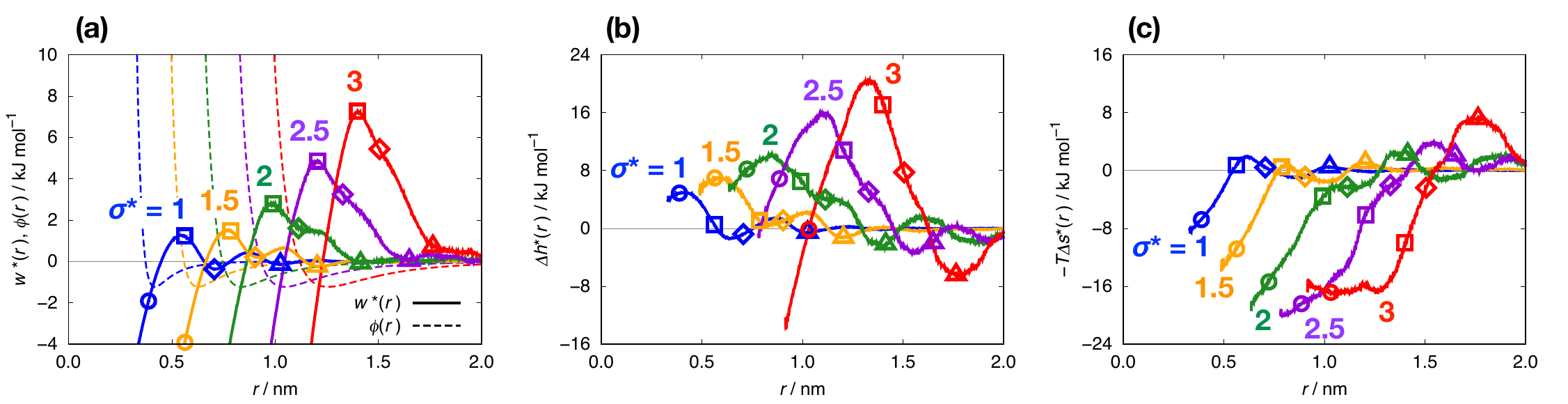}
\caption{
The water-mediated part $w^*(r)$ of the hydrophobic interaction and 
the enthalpic and entropic contributions, $\Delta h^*(r)$ and $-T\Delta s^*(r)$, to $w^*(r)$ 
for the solutes with the diameters $\sigma^* = 1, 1.5, 2, 2.5, {\mathrm{\ and \ }}3$ at $T=300{\mathrm{\ K}}$: 
(a) $w^*(r)$ (solid curve) and the direct pair potential $\phi(r)$ (dotted curve); 
(b) $\Delta h^*(r)$; and (c) $-T\Delta s^*(r)$. 
The open circles, squares, diamonds, and triangles indicate the separation distances of the 
first minimum, the first maximum, the second minimum 
(monolayer-separated minimum), and the third minimum (bilayer-separated minimum) of $w(r)$.
}
\label{fig_enthalpy_and_entropy}
\end{figure*}

\subsection{Enthalpic and entropic contributions to $w^*(r)$ and their solute-size dependences}

Next, we examine how the enthalpic and entropic contributions to $w^*(r)$ change with the solute size. 
Fig.~\ref{fig_enthalpy_and_entropy} shows curves of $w^*(r)$, $\Delta h^*(r)$, 
and $\Delta s^*(r)$ for $\sigma^*=$1, 1.5, 2, 2.5, and 3 at $T=300{\mathrm{\ K}}$. The circle, square, diamond, and triangle on each curve indicates 
the distances of the first minimum (contact minimum), the first maximum, the second minimum (monolayer-separated minimum), and the third minimum (bilayer-separated minimum) of the potential $w(r)$ of mean force. 

One can see in Fig.~\ref{fig_enthalpy_and_entropy}(a) that with increasing $\sigma$, the free-energy barrier of $w^*(r)$,  which corresponds to the first maximum of $w(r)$, rises sharply from 1~kJ/mol for $\sigma^*=1$ to 7~kJ/mol for $\sigma^*=3$. At the same time 
the local minimum of $w^*(r)$, which corresponds to the second minimum of $w(r)$, also goes up with $\sigma^*$ and disappears at $\sigma^*=2.5$. 
Comparing Fig.~\ref{fig_enthalpy_and_entropy}(b) with Fig.~\ref{fig_enthalpy_and_entropy}(a), 
one notices that the major peak of $\Delta h^*(r)$ corresponds
to that in $w^*(r)$ at a distance close to the first maximum of $w(r)$. 
Thus, the remarkable evolution of the free-energy barrier with increasing solute size is due to the increase in the unfavorable solvation enthalpy.
At the solute-solute contact distance, 
$\Delta h^*(r)$ is positive for $\sigma^*$ up to 2.5 but nearly zero for $\sigma^*=3$ [open circles in Fig.~\ref{fig_enthalpy_and_entropy}(b)] while 
$-T\Delta s^*(r)$ is always negative and the magnitude is much greater than $\Delta h^*(r)$ [open circles in Fig.~\ref{fig_enthalpy_and_entropy}(c)]. 
The minimum of $w(r)$ at the solute-solute contact distance 
goes down with increasing solute diameter because the entropic contribution ($-T\Delta s^*(r)$) becomes more negative. This statement is valid for $\sigma^*\le 2.5$; we will discuss a different cause for $\sigma^*=3$ and beyond.

\begin{figure*}[ht]
\centering
\includegraphics[scale=0.43]{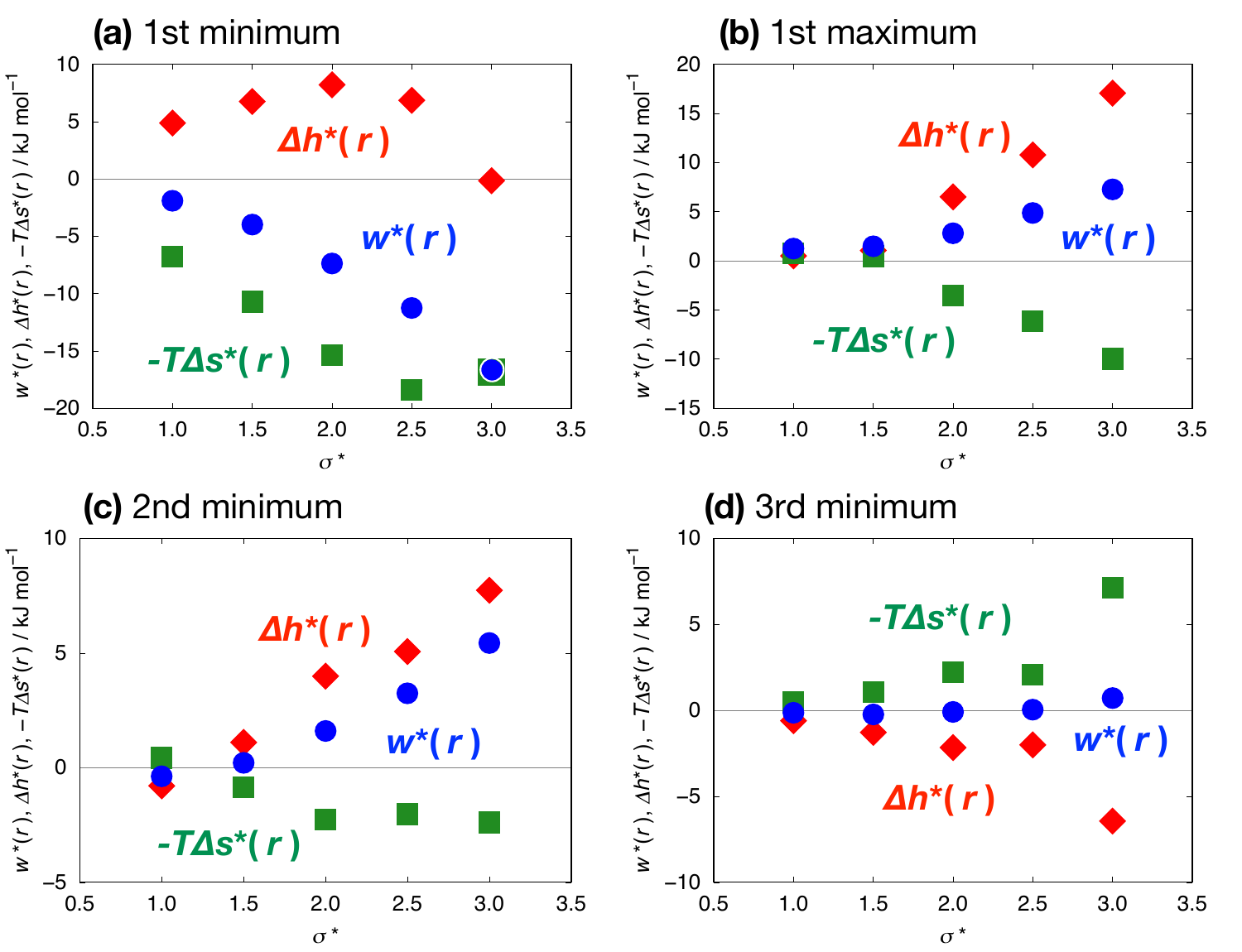}
\caption{
The solute size dependence of $w^*(r)$, $\Delta h^*(r)$, and $-T\Delta s^*(r)$ at $r$ of (a) the first minimum, (b) the first maximum, (c) the second minimum, and (d) the third minimum of $w(r)$.
The solute diameter $\sigma^*$ ranges from 1 to 3.
Blue circles, red diamonds, and green squares represent $w^*(r)$, $\Delta h^*(r)$, and $-T\Delta s^*(r)$, respectively.
}
\label{fig_peak_value}
\end{figure*}

Plotted in Fig.~\ref{fig_peak_value} are values of $w^*(r)$, $\Delta h^*(r)$, and $-T\Delta s^*(r)$ at fixed distances $r$ as functions of $\sigma^*$. 
At the solute-solute contact distance [Fig.~\ref{fig_peak_value}(a)],  
$w^*(r) (<0)$ decreases monotonically with $\sigma^*$; $-T\Delta s^*(r)$ decreases up to $\sigma^*=2.5$ and  turns to increase slightly; and $\Delta h^*(r)$ is maximal at around $\sigma^*=2$. 

{
For the methane-sized LJ particles in contact with each other, the magnitude of the favorable entropy change $\Delta s^*(r)/k~(\simeq 2.72)$ exceeds that of the unfavorable entropy change $\Delta h^*(r)/kT~(\simeq 1.95)$ to such an extent that the water-induced pair potential $w^*(r)/kT~(\simeq -0.77)$ is slightly negative. On the other hand 
for the C$_{60}$-sized LJ particles in the contact configuration, $\Delta h^*(r)/kT~(\simeq -0.06)$ is close to zero, and thus 
$w^*(r)/kT~(\simeq -6.69)$ is essentially equal to $-\Delta s^*(r)/k$.
}

At distances corresponding to the first maximum and second minimum of $w(r)$ 
[Figs.~\ref{fig_peak_value}(b) and~(c)], for the most part, 
$w^*(r)$ and $\Delta h^*(r)$, both positive, increase with $\sigma^*$
while $-T\Delta s^*(r)$ ($<0$) decreases with $\sigma^*$. 
At the distance of the third minimum of $w(r)$ or 
the bilayer-separated distance [Fig.~\ref{fig_peak_value}(d)],  
$w^*(r)$ is near zero in the whole range of $\sigma^*$. However, 
$-T\Delta s^*(r)$ ($>0$) increases with $\sigma^*$ 
while $\Delta h^*(r)$ ($<0$) decreases. 
For the largest solutes of $\sigma^*=3$, therefore, there is a large cancellation of 
$-T\Delta s^*(r)=7.14{\mathrm{\ kJ \ mol^{-1}}}$ and $\Delta h^*(r)=-6.42{\mathrm{\ kJ \ mol^{-1}}}$ resulting in $w^*(r)=0.72{\mathrm{\ kJ \ mol^{-1}}}$.

Having found the solvation-enthalpy and solvation-entropy contributions to $w^*(r)$, we now extract microscopic information from the spatial distributions of the potential energy $E_{\mathrm{ww}}$
and the number $N_{\mathrm{HB}}$ of hydrogen bonds, both per water molecule, around the pair of the hydrophobic solute particles in each system.

\begin{figure*}[ht]
\centering
\includegraphics[scale=0.094]{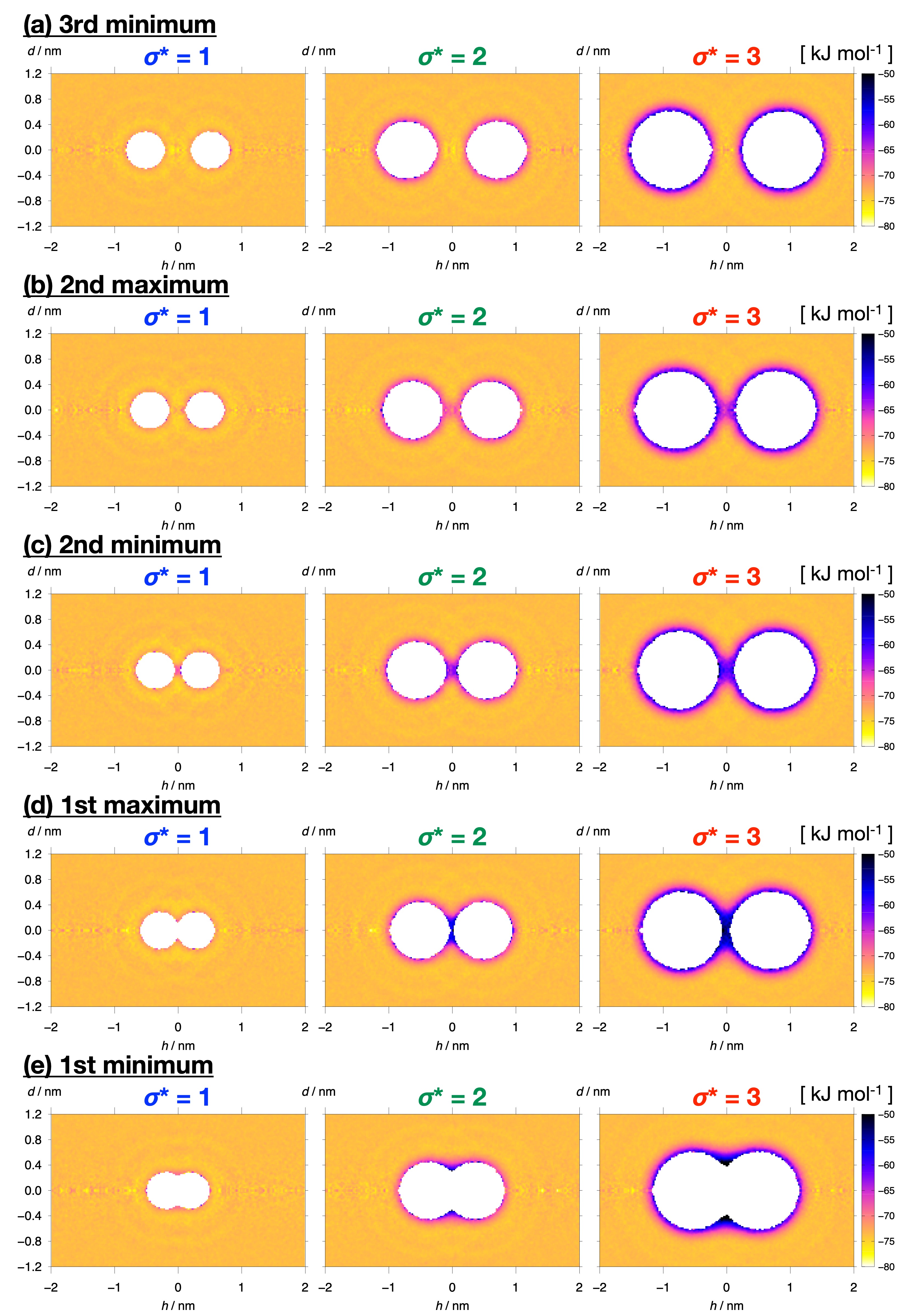}
\caption{
Spatial distributions of the potential energy $E_{\mathrm{ww}}$ of a water molecule 
due to the water-water interaction around the fixed pair of solute particles at $T=300{\mathrm{\ K}}$.
Two solutes with LJ diameters $\sigma^*=$ 1(left), 2(middle), and 3(right) are displayed in white.
The solute-solute separation distances are fixed to those of (a) the third minimum, (b) the second maximum, (c) the second minimum, (d) the first maximum, and (e) the first minimum of $w(r)$ for each solute (see Table~\ref{table_2}).
$E_{\mathrm{ww}}$ of the bulk water is $-73.7{\mathrm{\ kJ \ mol^{-1}}}$.
}
\label{fig_energy}
\end{figure*}

\subsection{Distributions of water-water interaction energy and the number of hydrogen bonds around pairs of solute particles}

The distribution $E_{\mathrm{ww}}(h,d)$ is the spatial distribution of the potential 
energy for a single water molecule due to water-water pair interactions within a cut-off distance of 0.35~nm. Figs.~\ref{fig_energy}(a)--(e) show 3$\times$5 distributions: the three sized solutes  ($\sigma^*=$ 1, 2, and 3) and the five distances $r$. 
When $r$ is of the third minimum of $w(r)$ [Fig.~\ref{fig_energy}(a)], the energy distribution in the vicinity of one spherical particle 
is unaffected by the other particle. For $r$ equal to or smaller than that of the second maximum of $w(r)$ [Figs.~\ref{fig_energy}(b)--(e)],  $E_\text{ww}$ in between the solute particles and in the neighborhood of the intersection circle of the two first-solvation shells around the two particles is higher than $E_\text{ww}$ in the other region.

Comparing three heat maps in each row of Fig.~\ref{fig_energy}, one finds that 
the larger the solute particles, $E_\text{ww}$ in the vicinity of each particle
is higher than $E_\text{ww}$ in the bulk. 
In Fig.~\ref{fig_energy}(a), for example, 
$E_{\mathrm{ww}}$ in the first solvation shells of the solute particles
of $\sigma^*=1$ is almost the same as the bulk value $-73.7{\mathrm{\ kJ \ mol^{-1}}}$, but the corresponding $E_{\mathrm{ww}}$ is $-65{\mathrm{\ kJ \ mol^{-1}}}$ and $-60{\mathrm{\ kJ \ mol^{-1}}}$ for the larger solute particles of diameters $\sigma^*=$~2 and 3, respectively. Water molecules in the first solvation shell are little perturbed 
by the methane-size solute but are more frustrated by the larger solutes. 
We will see the equivalent picture in terms of the hydrogen bonding.
\par

Except for $\sigma^*=1$, $E_{\mathrm{ww}}$ between the two spherical solutes is higher than in bulk water when the solute-solute distance $r$ is of the second maximum of $w(r)$ or shorter: in Fig.~\ref{fig_energy}(b), $E_{\mathrm{ww}}(0,0)$ at the midpoint is higher by 9~kJ~mol$^{-1}$ and 14~kJ~mol$^{-1}$ than 
bulk water for $\sigma^*=$ 2 and 3, respectively, and $E_{\mathrm{ww}}(0,0)$ increases as $r$ decreases further  [Figs.~\ref{fig_energy}(c) and (d)].

\begin{figure*}[ht]
\centering
\includegraphics[scale=0.094]{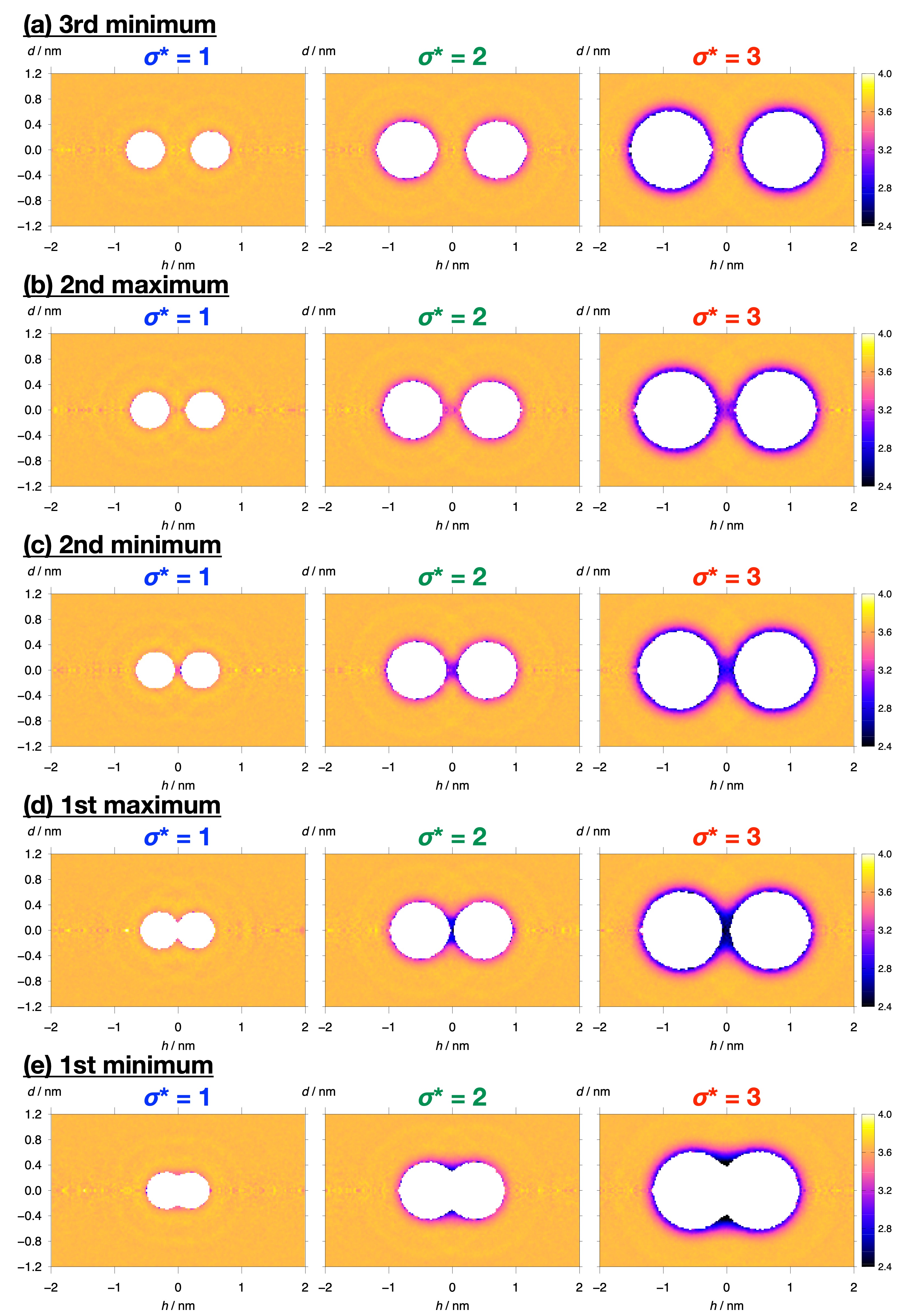}
\caption{
The numbers $N_{\mathrm{HB}}$ of hydrogen bonds between water molecules per one molecule at $T=300{\mathrm{\ K}}$.
Two solutes with LJ diameters $\sigma^*=$ 1(left), 2(middle), and 3(right) are displayed in white.
The solute-solute separations $R$ are fixed to $r$ of (a) the third minimum, (b) the second maximum, (c) the second minimum, (d) the first maximum, and (e) the first minimum of $w(r)$ for each solute (see Table~\ref{table_2}).
$N_{\mathrm{HB}}$ of the bulk water is 3.66.
}
\label{fig_hydrogen_bond}
\end{figure*}

The spatial distribution of the average number $N_{\mathrm{HB}}$ of hydrogen bonds per water molecule are shown in Fig.~\ref{fig_hydrogen_bond}. 
At the bilayer-separated distance, i.e., $r$ of the third minimum of $w(r)$ [Fig.~\ref{fig_hydrogen_bond}(a)], $N_{\mathrm{HB}}$ in the solvation shell of the small solute ($\sigma^*=1$) is almost the same as $N_{\mathrm{HB}}=3.66$ in the bulk water  while $N_{\mathrm{HB}}\approx3.4$ and 3.0 in the solvation shells of the solutes with $\sigma^*=2$ and 3, respectively. 
In accordance with what we saw for $E_{\rm ww}$, we notice that the hydrogen-bond network is hardly disturbed by the methane-sized solutes whereas it is strongly destabilised by the large particles. This result illustrates the difference in the hydrophobic hydration of small and large solute molecules---the idea expressed by Stillinger~\cite{stillinger1973} and  theoretically confirmed by Chandler et al.~\cite{lum1999, chandler2005nature}

When the distance between two particles is of the second
maximum of $w(r)$ [Fig.~\ref{fig_hydrogen_bond}(b)], there appears the effect of confinement on water: the effect is nearly absent for $\sigma^*=1$, slightly visible for $\sigma^*=2$ and clearly notable for $\sigma^*=3$. As the distance reduces to that of the second minimum of $w(r)$ [Fig.~\ref{fig_hydrogen_bond}(c)] and further down to that of the first maximum of $w(r)$ [Fig.~\ref{fig_hydrogen_bond}(d)], $N_{\mathrm{HB}}$ in the region between two particles becomes lower and lower. When the two particles of $\sigma^*=3$ are in contact with each other [Fig.~\ref{fig_hydrogen_bond}(e)], $N_{\mathrm{HB}}$ at the intersection of the two solvation shells [the black regions at the cusps in the figure] is close to 2.4 as compared to the bulk value 3.66. 
The set of the results for $N_{\mathrm{HB}}$ is all consistent with that for $E_{\mathrm{ww}}$.

Earlier studies showed that the structure of water confined between hydrophobic planar walls changes from bilayer-like to monolayer-like structure as two walls approach each other.~\cite{koga2002jcp,koga2005jcp,engstler2018jpcb,leoni2021ACSNANO}
Figs.~\ref{fig_distribution},~\ref{fig_energy}, and~\ref{fig_hydrogen_bond} show similar trends: 
When the solute-solute separation is of the third minimum of $w(r)$, the bilayer-like structure is present in between the two particles of $\sigma^*=3$. At the separation distance of the second minimum of $w(r)$, confined water assumes the monolayer-like structure. It is noteworthy that $N_{\mathrm{HB}}\simeq 3.6$ in the bilayer-like water is only slightly less than that in bulk water [the right panel in Fig.~\ref{fig_hydrogen_bond}(a)] while $N_{\mathrm{HB}}\simeq 2.9$ in the monolayer-like structure is much less that the bulk value [the right panel in Fig.~\ref{fig_hydrogen_bond}(c)]. This result is consistent with the earlier MD simulation result that the melting temperature of bilayer ice is much higher than that of monolayer ice:~\cite{koga2005jcp} 
water molecules can form four-coordinated network structure without substantial energetic penalty if there are two molecular layers or more in confining geometry.

\subsection{Molecular level picture of the difference in 
hydrophobic interactions of small and large particles}

\revise{Here we discuss, at the molecular level, the difference in hydrophobic interactions of small and large particles in the contact-minimum configuration.
}

\begin{figure*}[ht]
\centering
\includegraphics[scale=0.50]{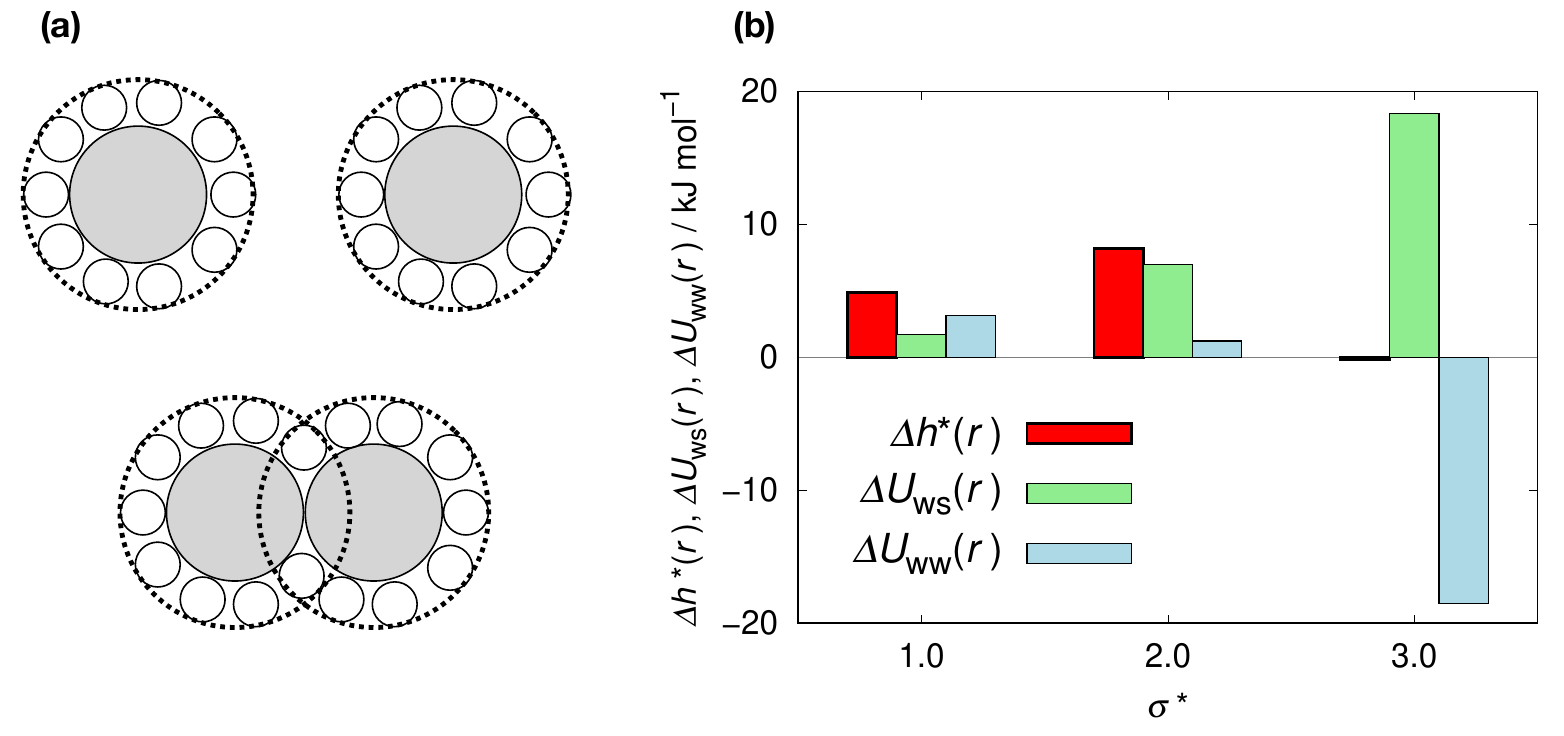}
\caption{Changes in the solvation shells and the solvation enthalpy when two solute particles are brought into contact. (a) Schematic picture of the first solvation shells around a pair of large solute particles in the separated and contact configuration. The water molecules in the solvation shells have $E_{\rm ww}$ higher than in bulk (Fig.~\ref{fig_energy}), $N_{\rm HB}$ less than in bulk (Fig.~\ref{fig_hydrogen_bond}), but their solute-water interaction energy is lower than in bulk. (b) the solvation enthalpy change $\Delta h^*(r)$ and its components $\Delta U_{\rm ww}(r)$ and $\Delta U_{\rm ws}(r)$.}
\label{fig_uws_uww}
\end{figure*}

\revise{
First, let us consider the positive entropy change $\Delta s^*$ in the contact-minimum configuration [Fig.~\ref{fig_peak_value}(a)]. The positive $\Delta s^*$ cannot be explained by an increase in the rotational entropy of water molecules, because the hydrogen bond number per molecule is lower in the first hydration shell than in the bulk water. This is due to an increase in the translational entropy of solvent molecules or, equivalently, the excluded volume effect, as shown schematically in Fig.~\ref{fig_uws_uww}(a). Since the excluded volume effect is present in all liquids, one observes that $\Delta s^* > 0$ for hard sphere solutes in a hard sphere solvent, LJ solutes in an LJ solvent, and hydrophobic solutes in water. A subtle issue is the solute size dependence of $\Delta s^*$ as shown in Fig.~\ref{fig_peak_value}(a): $\Delta s^*$ increases with the solute diameter and peaks out at about $\sigma^*=2.5$. 
The fact that $\Delta s^*$ does not increase monotonically with solute diameter may be due to a unique property of hydrophobic solutes in water. }

\revise{
Second, we recall that $\Delta h^*(r)=h^*(r) - h^*(\infty)$,  
where $h^*(r)$ is the solvation enthalpy of the pair of solute particles
and is essentially the excess of the configurational energy $U$ of 
the system with the pair over that without, for the pressure-volume term is negligible. 
The configurational energy $U$ consists of 
$U_{\rm ww}$ the sum of the pair interactions between water molecules and 
$U_{\rm ws}$ the sum of the pair interactions between water molecules and the solute pair. Note that 
\begin{equation}
    U_{\rm ww}(r)=\frac{\rho}{2}\int E_{\rm ww}(h,d; r) g_{\rm cyl}(h,d; r) d\tau,
\end{equation}
where $g_{\rm cyl}(h,d; r)$ and $E_{\rm ww}(h,d; r)$ are respectively the normalized density distribution and the local water-water interaction energy displayed in Figs.~\ref{fig_distribution}~and~\ref{fig_energy}, and the integral with the volume element $d\tau$ is over the whole volume. 
The solvation enthalpy contribution to $w^*(r)$ is then
expressed as 
\begin{equation}
    \Delta h^*(r)=\Delta U_{\rm ww}(r) + \Delta U_{\rm ws}(r). 
\end{equation}
We obtained $\Delta U_{\rm ws}(r)$ from MD simulations for the contact-minimum configuration and the infinitely separated configuration, and then evaluated $\Delta U_{\rm ww}(r)$ as $\Delta h^*(r)-\Delta U_{\rm ws}(r)$. 
The results for the three sized LJ solutes in water are shown in Fig.~\ref{fig_uws_uww}(b).
}

\revise{
For the LJ particles in the contact configuration $\Delta U_{\rm ws}(r)$ is positive, because the number of water molecules in the first solvation shells is smaller when the two solute particles are in contact than 
when they are far apart (Fig.~\ref{fig_uws_uww}(a)), and increases monotonically with the particle diameter $\sigma^*$, because the larger the particle, the greater the decrease in the number of water molecules in the solvation shells. This behavior of $\Delta U_{\rm ws}(r)$ should be observed for any solvent species as long as there is an attractive force
between solute and solvent molecules. 
On the other hand, the particle size dependence of $\Delta U_{\rm ww}(r)$ reflects the hydrogen-bonding property of water.  For the methane-sized solutes, $\Delta U_{\rm ww}(r)$ is slightly positive because water molecules in the vicinity of solute particles can form stronger hydrogen bonds when the solutes are far apart than they are in contact. For the C$_{60}$-sized solutes, however, $\Delta U_{\rm ww}(r)$ is large and negative because water molecules form weaker hydrogen bonds in the first solvation shells than in bulk. The opposite solute size dependencies of $\Delta U_{\rm ww}(r)$ and $\Delta U_{\rm ws}(r)$ give rise to the non-monotonic solute size dependence of $\Delta h^*(r)$ as shown in 
Fig.~\ref{fig_peak_value}(a). For the fullerene-sized LJ solutes in the contact configuration, $\Delta h^*(r)$ is close to zero because the positive $\Delta U_{\rm ws}(r)$ and the negative $\Delta U_{\rm ww}(r)$ largely cancel each other out.}

\revise{
 In summary, for the fullerene-sized LJ solutes ($\sigma^*=3$), the water-induced potential $w^*(r)$ is essentially the entropy change due to the excluded volume effect. The enthalpy contribution is close to zero for the reason described above. On the other hand, for methane-sized LJ solutes ($\sigma^*=1$) in water near room temperature, the magnitude of the favorable entropy and that of the unfavorable enthalpy are nearly balanced, so that $w^*(r)$ is slightly negative. }

\subsection{Effective interactions between `water-repellent' particles}

\begin{figure}[ht]
\centering
\includegraphics[scale=0.43]{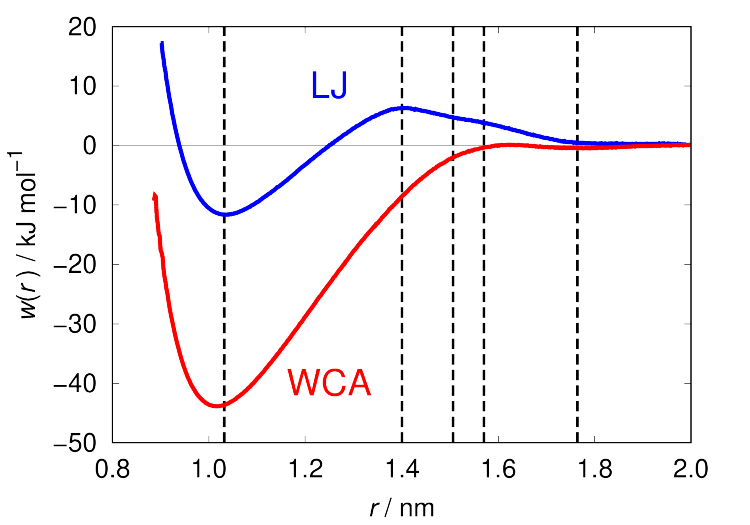}
\caption{
The potentials $w(r)$ of mean force for pairs of solutes with $\sigma^*=3$ in water at $T=300{\mathrm{\ K}}$, reported in the previous paper.~\cite{naito2024fd}
Blue and red curves are $w(r)$ for the solutes interacting with water molecules via the LJ potential and the repulsive WCA potential, respectively.
Black dot lines emphasize the distances $r$ of the first minimum and maximum, the second minimum and maximum, and the third minimum of $w(r)$ for the LJ particle with $\sigma^*=3$.
}
\label{fig_pmf_WCA}
\end{figure}

Up until now, we examined the solute-size effect on $w(r)$, $\Delta h^*(r)$, $-T\Delta s^*(r)$,
$E_{\rm ww}$, and $N_{\rm HB}$ under the condition that the LJ energy parameter 
for the solute-water interaction is fixed to the value for methane-water pairs. 
We shall now examine the effect of the weak solute-water attractive interaction on the solvent-induced potential $w^*(r)$. 
If the weak attraction is completely turned off, the potential of mean force curve changes significantly.  Compare in Fig.~\ref{fig_pmf_WCA}  the PMF curve obtained for solutes
interacting with water molecules via the repulsive part of the Weeks-Chandler-Andersen (WCA) potential (`water-repellent' solutes)  with the other for solutes interacting via the LJ potential, both types of solutes having the same diameter of $\sigma^*=3$.
First of all, the PMF at the solute-solute contact distance is significantly lower 
for the water-repellent solutes: the potential-well depth exceeds 40 kJ/mol.
Second, there exists no potential barrier between the contact distance and the monolayer- or bilayer-separated distance for the WCA solutes. 
\revise{
The reason why the C$_{60}$-sized WCA solutes attract each other much more strongly than the LJ solutes is that both $\Delta U_{\rm ws}$ and $\Delta U_{\rm ww}$ are negative and so $\Delta h^*(r)$ is large and negative; $\Delta s^*(r)$ is large and positive as it is for the LJ solutes. In summary, the large pairwise attractive interaction is driven by both the favorable entropy change and the favorable enthalpy change.
}

\revise{Note that for the all-atom model for C$_{60}$ molecules in water, the water-induced potential $w^*(r)$ at the contact-minimum distance is positive~\cite{makowski2010jpcb,li2005PRE,li2005JCP} in contrast to $w^*(r)$ being negative for the C$_{60}$ sized LJ and WCA particles. This is due to the strong water-C$_{60}$ attractive interaction.~\cite{li2005JCP}}

\begin{figure}[t]
\centering
\includegraphics[scale=0.09]{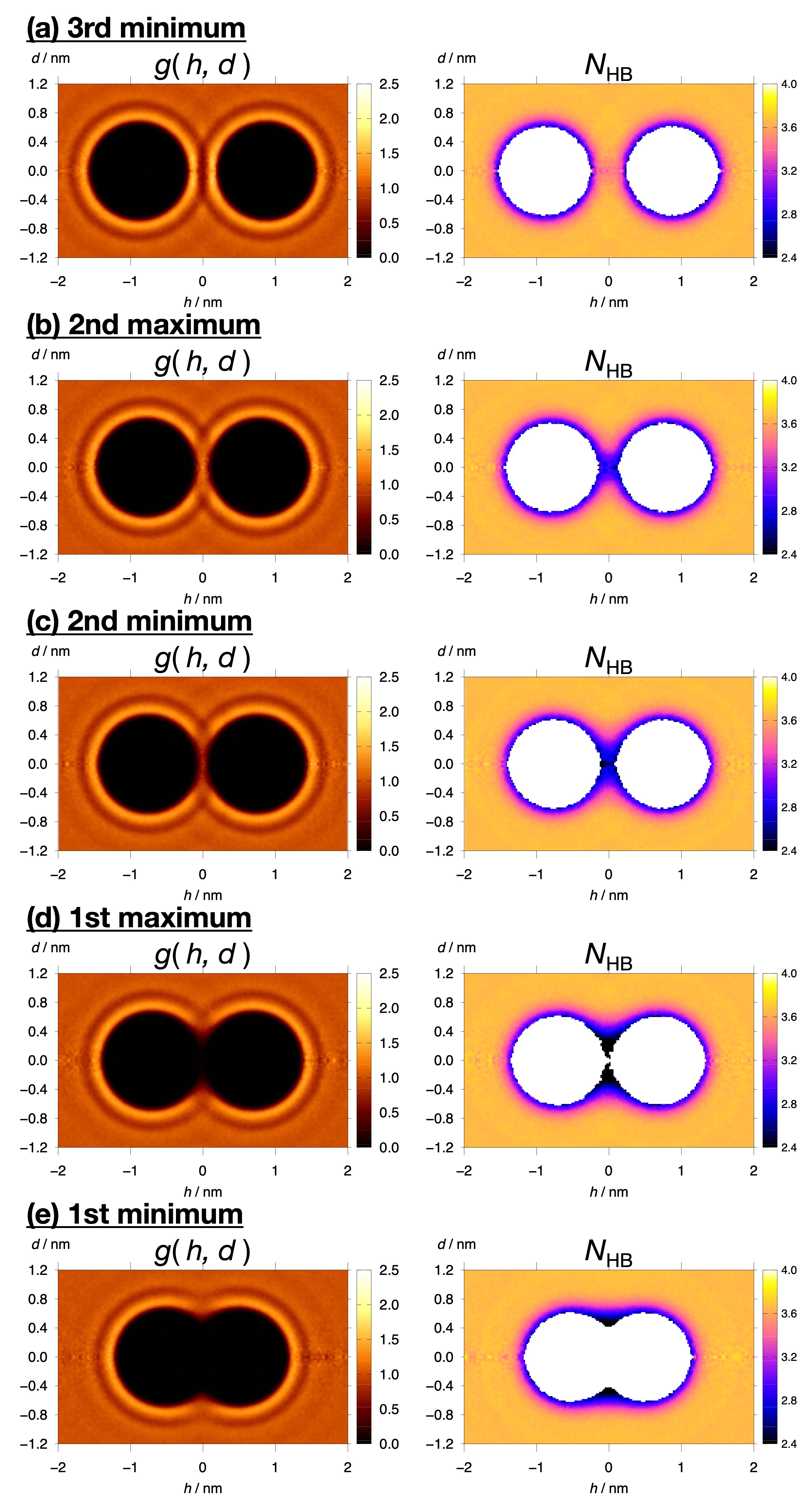}
\caption{Microscopic structures of water molecules near the two solute particles interacting with water molecules via the repulsive WCA potential at $T=300{\mathrm{\ K}}$. The solute LJ diameter is $\sigma^*=3$. (Left) The normalized two-dimensional distributions $g_\text{cyl}(h,d)$ of water molecules. (Right) The numbers $N_{\mathrm{HB}}$ of hydrogen bonds per one molecule. The solute-solute separations $R$ are $r$ of (a) the third minimum, (b) the second maximum, (c) the second minimum, (d) the first maximum, and (e) the first minimum of $w(r)$ for the LJ particle with $\sigma^*=3$ (see Table~\ref{table_2}).} \label{fig_heatmap_WCA}
\end{figure}

Fig.~\ref{fig_heatmap_WCA} shows $g_\text{cyl}(h,d)$ 
and $N_{\mathrm{HB}}(h,d)$ for the pair of WCA solutes with $\sigma^*=3$. 
When $r$ is of the third minimum of $w(r)$, or equivalently the bilayer-separated distance, 
$g_\text{cyl}$ in the immediate neighborhood of the solutes is approximately 1.4, which is slightly lower than that for the LJ solutes; however, 
$N_{\mathrm{HB}}$ is not significantly affected by the absence of the water-solute attractive interaction (Compare Fig.~\ref{fig_heatmap_WCA}(a) with Fig.~\ref{fig_hydrogen_bond}(a)). 
When $r$ is of the second maximum or of the second minimum of $w(r)$, 
water molecules between the WCA solutes have less hydrogen bonds than those between the LJ solutes. See Figs.~\ref{fig_heatmap_WCA}(b), (c) and Figs.~\ref{fig_hydrogen_bond}(b), (c). 
At the same time, the local density between the WCA solutes is smaller than that between the LJ solutes. 
When $r$ is of the first maximum, water molecules in between the WCA particles are completely depleted (Fig.~\ref{fig_heatmap_WCA}(d)) while 
those between LJ solutes are scarce but present (Fig.~\ref{fig_distribution}(d)).

We saw in Fig.~\ref{fig_pmf_WCA} that the free-energy barrier between the contact and solvent-separated configurations of two LJ particles in water disappears when the water-solute weak attractive interaction is completely turned off.
The microscopic mechanism is now clearly demonstrated by Fig.~\ref{fig_heatmap_WCA}(d): 
water molecules are so scarce or absent in the gap between the WCA particles that 
the two particles can approach each other without squeezing out water.

\section{Concluding Remarks}

\noindent
{We have observed that the nature of the hydrophobic interaction changes as the diameter of the solute increases from that of methane to that of C$_{60}$.
For the methane-sized solutes in the contact minimum configuration, the favorable entropy change $\Delta s^*(r)/k~(>0)$ and the unfavorable enthalpy change $\Delta h^*(r)/kT~(>0)$ largely cancel each other out to make the water-induced potential $w^*(r)/kT$ slightly negative. On the other hand for the C$_{60}$-sized solutes 
the water-induced potential $w^*(r)/kT$ is large and negative, entirely due to the favorable entropy change (Fig.~\ref{fig_peak_value}(a)).}
To focus on the solute-size effect, the LJ energy parameters for the solute-solute and solute-water pair potentials have been fixed to those for methane-methane and methane-water pairs. We demonstrated that $w^*(r)$ is very sensitive to the strength of the solute-solvent attractive interaction (Fig.~\ref{fig_pmf_WCA}). 

For small hydrophobic solutes such as methane, 
the potential $w(r)$ of mean force has the well-defined minima corresponding to  
the contact, monolayer-separated, bilayer-separated configurations. 
As the size of solute particles increases, the first minimum decreases monotonically and 
the PMF curve in the range between the contact minimum and the bilayer-separated minimum goes up and eventually bears no local extrema in that range. The consequence is the single free-energy barrier between the contact and bilayer-separated configurations. 

The decrement of the contact minimum in $w(r)$ 
with increasing solute diameter is accompanied by
the increasing favorable solvation entropy term $T\Delta s^*(r)$
which exceeds the increasing unfavorable solvation enthalpy term $\Delta h^*(r)$. The increase of the {\it excess} solvation entropy of the contact pair over that of the infinitely separated pair 
has been confirmed for both aqueous solutions and simple liquids
based on simulation and the theory of liquids.~\cite{lum1999,hummer1996pnas,sumi2007jcp,graziano2009jpcb,graziano2010cpl,graziano2017cpl,zangi2011jpcb,BenAmotz2015jpcl,naito2022jcp,naito2024fd}. 
The essential feature of the solute size dependence of $T\Delta s^*(r)$ has already been captured at the level of the Asakura-Oosawa excluded-volume effect,~\cite{asakura1954jcp, naito2024fd} and should therefore be considered as a common nature of solvent-induced pair interactions.
Whether the entropic term is the major factor in determining the solute-size dependence of the contact minimum in $w(r)$, however, depends on 
details of the solute-solvent and the solvent-solvent interactions.~\cite{BenAmotz2015jpcl} 


The development of the free-energy barrier between the contact pair
and the bilayer-separated pair with increasing size of the solute
is due to the increasing unfavorable solvation enthalpy contribution.
The free energy barrier disappears when the solute-water attractive interaction is replaced by a purely repulsive force (Fig.~\ref{fig_pmf_WCA}). We also know that in simple liquids the solvent-separated minima do not disappear with increasing solute size.~\cite{naito2024fd}  
A deeper insight on the free energy barrier would be gained by evaluating the enthalpic and entropic contributions to $w(r)$ for different sized solutes in simple liquids.

{
Two notable features of the solvation of pairs of hydrophobic solutes are derived (Figs.~\ref{fig_distribution},~\ref{fig_energy}, and~\ref{fig_hydrogen_bond}). First, methane molecules hardly perturb the hydrogen bond network of water while C$_{60}$-sized hydrophobic particles cause the surrounding water molecules to have fewer hydrogen bonds than in the bulk (Fig.~\ref{fig_hydrogen_bond}). For the C$_{60}$-sized solutes in the contact configuration, the water-water potential energy change $\Delta U_{\rm ww}(r)$ is large and negative, the water-solute potential energy change $\Delta U_{\rm ws}(r)$ is large and positive, and consequently the enthalpy change $\Delta h^*(r)$ is close to zero. That is, the hydrophobic attraction between the C$_{60}$-sized solutes is driven by the solvation entropy change $T\Delta s^*(r)$.} Second, the hydrogen bond network between two large hydrophobic solutes is much less perturbed when the two solutes are in the bilayer-separated configuration. That is, the signature of the solvation force between two macroscopic hydrophobic surfaces~\cite{koga2002jcp,koga2005jcp,engstler2018jpcb,leoni2021ACSNANO,li2005JCP} is already anticipated in the case of C$_{60}$-sized solutes. 

Finally, the power-law behavior [Eq.~(\ref{eq_B_sigma_power})] 
of the osmotic second virial coefficient $B$ has been found at 360~K,
affirming the robustness of the previously reported results.~\cite{naito2022jcp, naito2024fd} 

\begin{acknowledgments}
This work was supported 
by KAKENHI (Grant Numbers JP18KK0151 and JP20H02696) and 
by JST, the establishment of university fellowships towards the creations of science technology innovation (Grant Number JPMJFS2128).
Part of the computation was performed using Research Center for Computational Science, Okazaki, Japan (Project: 22-IMS-C124 and 23-IMS-C112).
\end{acknowledgments}

\bibliography{reference}

\end{document}